\documentstyle[12pt]{article}

\newcommand{\be}{\begin{equation}}
\newcommand{\ee}{\end{equation}}

\newcommand{\bea}{\begin{eqnarray}}
\newcommand{\eea}{\end{eqnarray}}

\begin{document}
\title{{\bf Pion and thermal photon spectra as a possible signal for a phase
transition}}
\author{
{\bf A.~Dumitru, U.~Katscher, J.A.~Maruhn, H.~St\"ocker, W.~Greiner}
\\[0.3cm]
{\small Institut f\"ur Theoretische Physik der J.W.Goethe Universit\"at}\\
{\small Postfach 111932, D-60054 Frankfurt a.M., Germany}\\
\\[0.5cm]
{\bf D.H.~Rischke}
\\[0.3cm]
{\small Physics Department, Pupin Laboratories, Columbia University}\\
{\small New York, NY 10027, USA}
}
\date{\today}
\maketitle
\newpage
\begin{abstract}
We calculate thermal photon and neutral pion
spectra in ultrarelativistic
heavy-ion collisions in the framework of three-fluid hydrodynamics.
Both
spectra are quite sensitive to the equation of state used. In particular,
within our model, recent data for $S+Au$ at $200$ AGeV
can only be understood if a scenario with a phase transition
(possibly to a quark-gluon plasma) is assumed. Results for
$Au+Au$ at $11$ AGeV and $Pb+Pb$ at $160$ AGeV are also presented.
\end{abstract}

\section{Introduction}
One of the most important goals of todays heavy-ion physics is the search
for the quark-gluon plasma (QGP), a phase of deconfined quark and gluon
matter which may be formed at high energy densities \cite{QGP}. If the plasma
is created in a heavy-ion collision, it will emit lots of particles which
may serve as `probes` of this novel phase of nuclear matter.
Electromagnetic probes, like real or virtual
photons, are of outstanding interest since they are not subject to strong
interactions and thus their mean free path is large enough to leave the
hot and dense reaction zone and carry information about its
properties to the detector \cite{em,KLS}.

Recently, the first (preliminary) single photon spectra in $S+Au$ collisions
at $200$ AGeV have been
presented by the WA80 group \cite{WA80}. After subtraction
of photons from $\pi^0$ and $\eta$ decays, data
seem to be in agreement with the spectrum of thermal radiation from a hot
hadronic and quark-gluon matter
source.
This was already observed in refs.\ \cite{Sriv,Neum}.
However, these calculations are based on assumptions for the
dynamical evolution of the system which are too simplified to allow for
reliable conclusions. In particular,\\
(a) in both references (longitudinal)
boost-invariant hydrodynamics \cite{Bj} was used.
This may be appropriate at collider energies but certainly not for
$E_{Lab}\le 200$ AGeV, where a considerable amount of stopping is
observed \cite{Matt,Hofm}, especially for heavy systems like $Pb+Pb$.
Therefore, we will solve the full relativistic
hydrodynamic equations of motion in (3+1) space-time
dimensions.\\
(b) In refs.\ \cite{Sriv, Neum} only the expansion stage of the collision
was considered.
The time $\tau_i$ (where expansion starts) is a free parameter which
may be related to the initial
temperature $T_i=T(\tau_i)$ by uncertainty-relation
arguments, and, assuming entropy conservation, to the final pion
multiplicity \cite{KMcLS,Sriv2}.
On the other hand, in our calculation the compressional stage of the collision
is consistently treated
and thus no such parameters appear. However,
if one-fluid hydrodynamic models are used,
the central energy density (at the time of maximal compression)
comes out much too large (not far from
the limit given by the Rankine-Hugoniot-Taub equation \cite{Taub}).
This is due to the assumption
of instantaneous local thermodynamic equilibrium and
presents one of the major problems of applying one-fluid hydrodynamics to
the early stage of
ultrarelativistic heavy-ion collisions. To solve this problem, we use
a three-fluid hydrodynamic model, as described below.\\
(c) In calculating photon production rates from a QGP or a hadron
gas, respectively,
one usually considers only the case of baryon-free matter which simplifies
the calculations considerably \cite{KLS}. However,
experiments \cite{Matt}, as well as dynamical models \cite{Hofm},
show considerable stopping in
nucleus-nucleus collisions up to $E_{Lab}= 200$ AGeV
(especially for heavy
systems), and there is little hope to create a baryon-free region, i.e., to
reach the Bjorken-limit\footnote{Note that the ratio of (net) baryons to pions
is considerably larger in the early stage of
the collision (where the large-$p_T$ photons resp.\ large-$M$ dileptons are
produced) than in the final state.}.
Therefore, in a one-fluid model, one
would have to account for finite baryon density effects. This
is not necessary in the three-fluid model, where separate fluids
for projectile, target, and produced particles are
used, since in this case the third fluid, which is by far the hottest and thus
gives the dominant contribution to the thermal radiation, is indeed
baryon-free.\\
(d) The photon spectrum measured by experiment is
dominated to 97\% by $\pi^0$ and $\eta$ decays \cite{WA80}.
Thus, before comparing calculated and measured thermal spectra
one first has to ensure that the dominant part of
the spectrum is reproduced by the dynamical model, i.e., that the underlying
hadron dynamics is consistent with experiment. To check this important
requirement, which is violated by boost-invariant hydrodynamics, we also
calculated the transverse momentum and rapidity spectra of
pions within our model.
The outline of the paper is as follows. In section 2 we present the
three-fluid model as used here and compare calculated pion spectra with
experimental data. We shall see that agreement is found only if a phase
transition (possibly to a quark-gluon plasma) is assumed at high
energy densities. Section 3 contains a brief discussion of the thermal photon
rate from quark and hadron matter sources, respectively. In section 4
we calculate photon spectra and compare them to available experimental data.
As was the case for pion observables, data seem to favour a scenario
with a phase transition. Section 5 concludes this work with a summary
of our results. We use natural units $\hbar=c=k_B=1$.

\section{The three-fluid model}
The original one-fluid hydrodynamic model \cite{mar85} represents,
besides microscopic models \cite{Micro}, one possibility to describe
the dynamics of heavy-ion collisions.
However, as discussed above, it assumes
local thermodynamic equilibrium and thus is
inappropriate to describe the initial stage of
ultrarelativistic collisions, at least for
$E_{Lab}\ge 10$ AGeV.
This problem is solved here by considering
more than one fluid \cite{ams00,mis00}.
The three-fluid model \cite{kat93} divides the particles involved in a reaction
into three separate fluids: the projectile nucleons, the target nucleons,
and the particles produced during the reaction.
Thermodynamic equilibrium is maintained only in each fluid separately
but not between the fluids. The fluids are able to penetrate
and decelerate while interacting mutually. This provides a means to treat
non-equilibrium effects in the initial stage of the collision.

The basic equations are
\bea \label{con1}
 \partial_\mu j^{\mu}_i & = & S_i \, , \\ \label{con2}
 \partial_\mu T^{\mu\nu}_i & = & S^\nu_i \, .
\eea
Here $j^\mu_i$ are the baryon density four-currents,
$T^{\mu\nu}_i$ the energy-momentum tensors,
and $S_i$, $S^\nu_i$ the
source terms which parametrize the interaction between the fluids.
The index $i=1,2,3$ labels the different fluids
(projectile, target, and produced particles).
Let $e_i$, $p_i$, $\rho_i$, and $U^\mu_i$
denote the local energy density, the pressure,
the local (net) baryon density, and the four-velocity, respectively,
of fluid $i$.
$j^\mu_i$ and $T^{\mu\nu}_i$
are then given by
\bea \label{jmu}
 j^{\mu}_i & = & \rho_i U^\mu_i \, , \\ \label{tmunu}
 T^{\mu\nu}_i & = & U^\mu_i U^\nu_i (e_i+p_i) -p_i g^{\mu\nu} \, .
\eea
If $S_i=S_i^\nu=0$, eq.~(\ref{con1}) represents baryon-charge conservation
and eq.~(\ref{con2}) energy-momentum conservation in fluid $i$.

Since the third fluid contains only particles produced during the reaction,
there is no net loss of baryons in projectile and target fluid, i.e.,
$S_1=S_2=\rho_3\equiv 0$, and
eq.~(\ref{con1}) does not need to be solved for the third fluid.
We assume chemical equilibrium in the third fluid
and thus the particle densities in that fluid
can be inferred from the energy density determined by eq.~(\ref{con2}).

The source terms $S^\nu_i$ can be split into interactions with each
of the other fluids
\be
 S^\nu_i = \sum _{j\neq i} s^\nu_{ij} \, .
\ee
$s^\nu_{ij}$ is supposed to be a superposition of
binary hadron collisions. This means
$s^\nu_{ij} = C_{ij} \delta p_{ij}^\nu$,
where $C_{ij}$ is the
rate of binary collisions and $\delta p_{ij}^\nu$
the average four-momentum loss of a particle in a binary collision.
The collision rate is given by
$C_{ij} = \rho_i \rho_j \sigma_{ij} v_{ij}$, where
$\sigma_{ij}$ is the total cross section of the free, binary collision,
$v_{ij}$ is the covariant relative velocity
$ v_{ij}^2 = (U_i^\mu U_{j,\mu})^2 -1$, and now $\rho_3$ stands for the
density of particles in the third fluid.
For the projectile-target interaction,
$\delta p_{12}^\nu$ can be
extracted from nucleon-nucleon data \cite{tan83}.
Since the third fluid is allowed to undergo a phase transition
to a quark-gluon plasma, $\delta p^\nu_{j3}$ ($j=1,2$) cannot be determined
experimentally for the interaction between the third fluid and the target
resp.\
projectile. For this ``rescattering'', we simply assume no energy exchange
and $50\%$ momentum loss in the center of mass system of the colliding
fluid elements.

The equation of state (EOS) of the target and projectile
fluids is that of an ideal nucleon gas
plus compression terms. We use a linear ansatz for the compression energy
with a compressibility of $250$ MeV
and a binding energy of $16$ MeV.

The EOS of the third fluid is that of an ideal gas of massive
$\pi$-, $\rho$-, $\omega$-, and $\eta$-mesons. At temperatures
$T\approx 100-250$ MeV it is not appropriate to use an equation of
state of an ideal pion gas, as done in ref.\ \cite{Neum}.
At $T_C=160$ MeV we allow for
a first order phase transition into a QGP.
For the QGP we then use the bag-model
EOS for (pointlike, massless, and noninteracting) $u$ and $d$ quarks.
The bag constant is chosen in such a way that the pressures of both phases
coincide at $T=T_C$.

Before presenting transverse momentum spectra of pions and photons, let us
first consider the rapidity distribution of negatively charged
hadrons in $O(200$ AGeV$)+Au$ and $S(200$ AGeV$)+S$,
which represents an additional test for our dynamical model, in that the
hadronic reaction dynamics is well described\footnote
{To our knowledge, no such data are published for $S+Au$.}.
We already pointed out that models assuming (strict) boost-invariance
fail this test. Fig.\ 1 shows
that data \cite{Rapd} are reproduced with sufficient accuracy. This is
no longer the case if no phase transition is allowed \cite{kat93}.
We also show a prediction for $Pb+Pb$.

One observes in fig.\ 2 that also the
calculated $\pi^0$ transverse momentum distribution
agrees well with the (preliminary)
reconstructed spectra of the WA80 group \cite{WA80}.
If, instead, no phase
transition is allowed, i.e., if we apply the hadronic EOS for all
energy densities, the pion flow is stronger and there are too many
pions at large $k_T$. The first scenario is obviously favored by the data.
In this figure we also present our results for $Pb(160$ AGeV$)+Pb$
collisions.

At this point we have established that our model reasonably
describes hadron dynamics and that pion spectra
are also reproduced correctly. Let us now turn to calculations of thermal
photon spectra.

\section{Thermal photon rate}
According to ref.\ \cite{KLS} the thermal photon production rate from an
equilibrated, baryon-free QGP is given (to first order in $\alpha$ and
$\alpha_S$) by
\be \label{rate}
E\frac{dR^\gamma}{d^3k} = \frac{5\alpha \alpha_S}{18\pi^2}
T^2 e^{-E/T} \ln \left( \frac{2.912E}{g^2 T} +1 \right)
\ee
where $E$ is the photon energy in the local rest frame of the QG-matter.
In the following calculations we fix $\alpha_S=g^2/4\pi=0.4$.
As shown in ref.\ \cite{KLS}, the rate for a
gas consisting of $\pi$-, $\rho$-, $\omega$-, and $\eta$-mesons
may also be parametrized by eq.\ (\ref{rate}). Other contributions, e.g.\ from
the $A_1$ meson \cite{Song}, as well as the effect of hadronic
formfactors \cite{KLS}, are neglected since they are of the
same magnitude as higher order corrections to eq.\ (\ref{rate}), which we
have also not taken into account.
We thus apply eq.\ (\ref{rate}) for both
phases of the third fluid. The contributions
from the first two fluids are negligible since (for the reactions considered
here) these fluids are much cooler. Also,
since they undergo a rapid longitudinal expansion,
they cool much faster than the third fluid.

\section{Results and discussion}
Our results are presented in figures 3-5 which show photon spectra for
central $Au+Au$ collisions at
$11$ AGeV, $S+Au$ at $200$ AGeV, and for $Pb+Pb$ at $160$ AGeV.
At the AGS, no pure QGP phase is created in our model.
However, a comparatively long-lived mixed phase does exist, and as a
consequence the thermal photon spectrum depends (at least for photons
with large transverse momentum $k_T\ge 1~GeV$) on whether
a phase transition to a QGP
happens or not. However, the thermal yield
at large transverse momentum is probably too low as compared
to the background of decay photons to be cleanly separated (for
$k_T\ge 1~GeV$ we estimate a ratio $\gamma_{thermal}/\pi^0 \le 1\%$).

In $S+Au$ collisions at the SPS, the third fluid reaches temperatures
up to
$T_{max}\approx 250$ MeV and thus a pure QGP phase does exist in our model.
For $Pb+Pb$, the maximum temperature is almost
the same but the space-time volume of the QGP is much larger.
Fig.\ 4 indicates that our scenario can only fit the WA80
data if a phase transition
is assumed; otherwise the slope and magnitude of the photon
spectrum is inconsistent with data, due to the fact that a hadronic equation
of state including only light mesons has less degrees of
freedom and is therefore hotter (at the same energy density).
Also, the pressure at high
energy densities is larger if no phase transition occurs and thus the
transverse flow is enhanced.
This is seen even better in $Pb+Pb$ collisions. We do also point out
that in our full $(3+1)$-dimensional calculation
the cooling of the system during the first few fm/c is slower
and the final transverse flow is stronger
as compared to boost-invariant hydrodynamics. In
$Pb+Pb$ collisions at the SPS this results in
a considerable suppression of high transverse momentum photons
in ref.\ \cite{Sriv2}.
However, our calculation is more realistic than that of refs.\
\cite{Sriv,Neum,Sriv2}
in that the initial conditions for the expansion of the third fluid are
self-consistently determined in the framework of the three-fluid model,
whereas in refs.\
\cite{Sriv,Neum,Sriv2} they are inferred from an uncertainty--relation
argument and the final pion multiplicity under the assumption of
entropy conservation. The latter assumption is questionable in view of the
well-established existence of entropy--creating hadronizing rarefaction
shock waves in the hydrodynamic expansion of matter undergoing a phase
transition \cite{Blaizot}. Moreover, our self-consistent calculation
does not impose the additional assumption of a longitudinal Bjorken-type
velocity profile, which is unrealistic at SPS energies.
Photon spectra for $Pb+Pb$ collisions might thus help to decide whether
longitudinally boost-invariant collision dynamics has to be
ruled out for SPS energies.

\section{Conclusions}
In conclusion, we have
presented an essentially parameter-free hydrodynamical
calculation of ultrarelativistic
heavy-ion collisions, established that hadronic observables are well
reproduced, and
shown that a scenario where an equilibrated QGP is
created strongly deviates from a purely hadronic scenario (with
light mesons only), e.g.\
in the thermal photon radiation (even if the QGP does not outshine the hot
hadronic gas) or the pion transverse momentum distribution.
Moreover, within our model, both of
these two independent observables are in agreement
with recently presented
data \cite{WA80} only if a phase transition to a QGP is allowed.
Nevertheless, future work should establish whether
these results cannot be reproduced with other
equations of state for the hadronic phase.
Indeed, our results are not sensitive on the exact form of the EOS, as long
as it shows a rapid increase of energy density in a narrow
temperature interval. This is sufficient to create a hydrodynamical flow
pattern and energy densities similar to those occuring in our
calculation. From hydrodynamics alone we can therefore not uniquely
specify the nature of the relevant degrees of freedom. For instance,
at vanishing baryon density the $\sigma-\omega$ model for nuclear
matter \cite{Theis}
exhibits an EOS very similar to ours.
Alternatively, one might
consider a Hagedorn gas \cite{Hage} with exponentially increasing
mass spectrum, which also reaches lower temperatures and pressures than
the gas of light mesons employed in our studies.
The thermal radiation from such matter, however, might be quite different
and, upon comparison with experiments, may give further clues with respect
to the nature of strongly interacting matter.
Furthermore, calculations within
microscopic, non-thermal models which do not incorporate a phase transition
are in progress.

\vspace*{1cm}
{\bf Acknowledgements:}
Helpful criticism, comments, and discussions with
H. Gutbrod, K.H. Kampert, and D.K. Srivastava are gratefully acknowledged.
D.H.R. thanks the Alexander v.\ Humboldt-Stiftung for
support under the Feodor-Lynen program and the Director, Office of Energy
Research, Division of Nuclear Physics of the Office of High Energy and
Nuclear Physics of the U.S.\ Dept.\ of Energy, for support under contract
No.\ DE-FG-02-93ER-40764.

\clearpage
\subsection*{Figure Captions}
{\bf Figure 1}
\\[0.2cm]
Rapidity distribution of negatively charged hadrons for central
$S+S$, $O+Au$, and $Pb+Pb$ collisions at the CERN-SPS, calculated within
the three-fluid model.
Data from Ref.\ \cite{Rapd}.
\\[1cm]
{\bf Figure 2}
\\[0.2cm]
Transverse momentum distribution of midrapidity (i.e.\ $y_{Lab}=3$)
neutral pions in central $S(200$ AGeV$)+Au$ collisions. The full curve
was calculated with and the dotted one without a phase transition.
The crosses and triangles show our results for $Pb+Pb$ at $160$ AGeV, divided
by $1000$.
\\[1cm]
{\bf Figure 3}
\\[0.2cm]
Thermal spectrum of midrapidity photons (i.e.\ $y_{Lab}=1.6$)
in central $Au(11$ AGeV$)+Au$ collisions, calculated within the three-fluid
model.
The full curve results when a phase transition is allowed, the dotted one
when it is not.
\\[1cm]
{\bf Figure 4}
\\[0.2cm]
Same as Fig.\ 3 but for $S(200$ AGeV$)+Au$ collisions ($y_{Lab}=3$).
\\[1cm]
{\bf Figure 5}
\\[0.2cm]
Same as Fig.\ 3 but for $Pb(160$ AGeV$)+Pb$ collisions ($y_{Lab}=3$).
\end{document}